\documentclass{PoS}

\title{Distance between configurations in MCMC simulations and the geometrical optimization of the tempering algorithms\footnote{Report No.: KUNS-2778}
}

\ShortTitle{Distance between configurations and the geometrical optimization}

\author{Masafumi Fukuma,$^a$ \speaker{Nobuyuki Matsumoto}$^a$ and Naoya Umeda$^b$\\
  \llap{$^a$}Department of Physics, Kyoto University\\
  Kyoto 606-8502, Japan\\
  \llap{$^b$}PrincewaterhouseCoopers Aarata LLC\\
  Otemachi Park Building, 1-1-1 Otemachi, Chiyoda-ku, Tokyo 100-0004, Japan\\
  E-mail:
  \email{fukuma@gauge.scphys.kyoto-u.ac.jp},
  \email{nobu.m@gauge.scphys.kyoto-u.ac.jp},
  \email{naoya.umeda1134@gmail.com}
}

\abstract{%
For a given Markov chain Monte Carlo (MCMC) algorithm, 
we define the distance between configurations 
that quantifies the difficulty of transitions. 
This distance enables us to investigate MCMC algorithms in a geometrical way,
and we investigate the geometry of the simulated tempering algorithm 
implemented for an extremely multimodal system with highly degenerate vacua. 
We show that the large scale geometry of the extended configuration space 
is given by an asymptotically anti-de Sitter metric, 
and argue in a simple, geometrical way 
that the tempering parameter should be best placed exponentially 
to acquire high acceptance rates for transitions in the extra dimension. 
We also discuss the geometrical optimization of 
the tempered Lefschetz thimble method, 
which is an algorithm towards solving the numerical sign problem. 
}

\FullConference{37th International Symposium on Lattice Field Theory - Lattice2019\\
  16-22 June 2019\\
  Wuhan, China}

\usepackage{mathrsfs,amsbsy,amssymb,latexsym,amsfonts,amsmath,amsthm}
\usepackage{graphicx,color}

\newcommand{\calM}{{\mathcal{M}}}
\newcommand{\calA}{{\mathcal{A}}}
\newcommand{\peq}{{p_\mathrm{eq}}}
\newcommand{\Peq}{{P_\mathrm{eq}}}
\newcommand{\bbW}{\mathbb{W}}

\begin{document}

\section{Introduction}
\label{sec:introduction}

In Markov chain Monte Carlo (MCMC) simulations, 
we often encounter a multimodal distribution, 
for which transitions between configurations 
around different modes are difficult. 
We introduced in \cite{FMU1} the {\it distance between configurations} 
to enumerate the difficulty of transitions. 

In this talk, we mainly consider the simulated tempering algorithm 
implemented for an extremely multimodal system with highly degenerate vacua. 
Our distance enables us to investigate the algorithm 
in a geometrical way as follows. 
We first define a metric on the extended configuration space, 
and show that it is given by an asymptotically anti-de Sitter (AdS) metric 
\cite{FMU1, FMU2}. 
We then show in a simple, geometrical way that 
the tempering parameter should be best placed exponentially 
to acquire high acceptance rates for transitions in the extra dimension. 
We further discuss the optimized form of the tempering parameter 
in the tempered Lefschetz thimble method (TLTM) 
\cite{Fukuma:2017fjq,Fukuma:2019wbv,FMU6}, 
which is an algorithm towards solving the numerical sign problem. 
This talk is based on work \cite{FMU1,FMU2,Fukuma:2019wbv}. 

\section{Definition of distance}
\label{sec:def_distance}

In this section, we briefly review the distance introduced in \cite{FMU1}. 
Let $\calM \equiv \{x\}$ be a configuration space, 
and $S(x)$ the action. 
Suppose that we are given an MCMC algorithm 
which generates a configuration $x$ from $y$ 
with the conditional probability $P(x|y)=(x|\hat{P} |y)$. 
We assume that it satisfies the detailed balance condition 
with respect to 
$\peq(x) \equiv (1/Z) e^{-S(x)} ~ 
\left( Z \equiv \int dx \, e^{-S(x)} \right)$. 
We further assume that 
the Markov chain satisfies suitable ergodic properties 
so that $\peq(x)$ is the unique equilibrium distribution. 

To define the distance, we consider the Markov chain in equilibrium. 
We denote by $\bbW_n$ the set of transition paths 
with $n$ steps in equilibrium, 
and by $\bbW_n(x,y)$, which is a subset of $\bbW_n$, 
the set of transition paths with $n$ steps 
which start from $y$ and end at $x$ in equilibrium. 
We define the {\it connectivity} between $x$ and $y$ 
by the fraction of the sizes of the two sets: 
\begin{align}
  f_n(x,y) \equiv \frac{|\bbW_n(x,y)|}{|\bbW_n|} = P_n(x|y)\peq(y) = f_n(y,x). 
\end{align}
Here $P_n(x|y)=(x|\hat{P}^n |y)$ is an $n$-step transition matrix. 
We further introduce the normalized connectivity as 
\begin{align}
  F_n(x,y) \equiv \frac{f_n(x,y)}{\sqrt {f_n(x,x)f_n(y,y)} }, 
\end{align}
with which we define the {\it distance} as follows: 
\begin{align}
  d_n(x,y) \equiv \sqrt{-2 \ln F_n(x,y) }. 
  \label{eq:def_distance}
\end{align}
It can be shown that this distance gives a universal form at large scales 
for algorithms that generate local moves 
in the configuration space \cite{FMU1}. 

As an example, let us first consider the action
$S(x) = (\beta/2)\sum_{\mu=1}^D x_\mu^2$, 
which gives a Gaussian distribution in equilibrium. 
The distance can be calculated analytically for the Langevin algorithm: 
\begin{align}
  d_n(x,y) = \frac{\beta}{2\sinh (\beta n \epsilon)}|x-y|^2, 
\end{align}
where $\epsilon$ is the increment of the fictitious time. 
We thus find that a flat and translation invariant metric 
is obtained for Gaussian distributions. 

As a second example, we consider the double-well action 
$S(x) = (\beta/2)(x^2-1)^2$, 
which gives a multimodal equilibrium distribution. 
We again use the Langevin algorithm to calculate the distance. 
By making use of a quantum mechanical argument and an instanton calculation, 
we find that the distance behaves for $\beta\gg 1$ as 
\begin{align}
 d_n(x,y) &= O(e^{-\beta n\epsilon/2}) 
 \quad (\mbox{when $x$, $y$ are around different modes}) 
\nonumber
\\
 d_n(x,y) &\propto \beta 
 \quad (\mbox{when $x$, $y$ are around the same mode}).
\end{align}
Therefore, we confirm that our distance 
quantifies the difficulty of transitions.

\section{Emergence of AdS geometry and the geometric optimization}
\label{sec:ads}

\subsection{Distance in the simulated tempering}
\label{sec:sim_temp}


The simulated tempering \cite{Marinari:1992qd} is an algorithm to 
speed up the relaxation to equilibrium. 
In this algorithm, we choose a parameter $\beta$ in the action 
(e.g.\ an overall coefficient) 
as the tempering parameter, 
and extend the configuration space in the $\beta$ direction: 
$\calM \rightarrow \calM\times\calA$, 
where $\calA \equiv \{\beta_0,\beta_1,\cdots,\beta_A\} 
= \{\beta_a\}_{a=0,\cdots,A}$ 
and $\beta_0$ is the original parameter of interest. 
We assume that $\{\beta_a\}$ are ordered 
as $ \beta_0 > \beta_1 \cdots > \beta_A $. 
We set up a Markov chain in the extended configuration space $\calM \times \calA$ 
in such a way that the global equilibrium distribution becomes 
$\Peq(x,\beta_a) \equiv w_a \exp[-S(x;\beta_a)]$. 
We choose the weights $\{w_a\}$ to be $w_a = 1/(A+1)Z_a$ $(Z_a \equiv \int dx \, e^{-S(x;\beta_a)})$. 
Expectation values are to be calculated by 
first realizing global equilibrium 
and then retrieving the subsample at $\beta_{a=0}$. 

For multimodal systems, 
transitions between configurations around different modes are difficult. 
This situation can get improved by extending the configuration space as above 
because then such
configuration can communicate easily by passing through 
the region with small $\beta_a$. 
The benefit due to tempering can be enumerated in terms of the distance. 
Table \ref{table:distance} shows 
the distance between two modes in the original configuration space 
with and without tempering for the double-well action. 
We see that the introduction of the simulated tempering 
drastically reduces the distance. 

\begin{table}[ht]
\centering
\begin{tabular}{|c|c|c|}
  \hline
  $n$ & without tempering & with tempering  \\
  \hline
  $10$ & $39.1$ & $26.5$ \\
  \hline
  $50$ & $19.2$ & $7.16$ \\
  \hline
  $100$ & $16.9$ & $4.35$ \\
  \hline
  $500$ & $13.2$ & $0.708$ \\
  \hline
  $1,000$ & $11.7$ & $0.106$ \\
  \hline
  $5,000$ & $8.46$ & $2.78\times 10^{-8}$ \\
  \hline
\end{tabular}\\
\caption{Comparison of the distance with and without tempering \cite{FMU1}.}
\label{table:distance}
\end{table}

\subsection{Emergence of AdS geometry}
\label{sec:emergence}

We hereafter consider an extremely multimodal system 
with highly degenerate vacua. 
As a typical example, we use the action 
$S(x;\beta) \equiv \beta\sum_{\mu=1}^D \left(1-\cos(2\pi x_\mu) \right)$. 

According to the definition of our distance, 
$d_n(x,y)$ is negligibly small when $x$ and $y$ lie around the same mode, 
while $d_n(x,y)$ is large when $x$ and $y$ are around different modes. 
Therefore, when we investigate the large-scale geometry of $\mathcal{M}$, 
we can identify configurations around the same mode as a single configuration. 
We write the coarse-grained configuration space thus obtained 
as $\bar{\calM}$ 
(for the cosine action, $\bar{\calM} = \mathbb{Z}^D$). 
We can similarly coarse-grain the extended configuration space 
when the simulated tempering is implemented. 
We write the extended, coarse-grained configuration space as 
$\bar{\calM} \times \calA$. 

We define the metric on 
$\bar{\calM} \times \calA = \{ X \equiv (x,\beta_a) \} $ in terms of the distance: 
\begin{align}
  ds^2 = g_{\mu\nu} dX^\mu dX^\nu =  d_n^2(X,X+dX), 
\end{align}
where $X$ and $X+dX$ denotes nearby points in $\bar{\calM} \times \calA$. 
It can be shown that this metric is an asymptotically AdS metric \cite{FMU2}. 
We here sketch the proof. 
We first note that the action is invariant under the lattice translation 
$x_\mu \rightarrow x_\mu + m ~ (m\in\mathbb{Z})$ 
and thus, the metric components are independent of $x$. 
Furthermore, since the action is also invariant under the reflection 
$x_\mu \rightarrow -x_\mu$, 
there is no off-diagonal components. 
Thus we deduce that the metric takes the following form: 
\begin{align}
  ds^2 = f(\beta) d\beta^2 + g(\beta)\sum_{\mu=1}^D dx_\mu^2. 
\end{align}
We are left with determining two functions $f(\beta)$, $g(\beta)$. 

We first consider $g(\beta)$. 
Since transitions in the $x$ direction are difficult for larger $\beta$, 
$g(\beta)$ should be an increasing function of $\beta$ 
at least when $\beta$ is large. 
We here assume that the leading dependence on $\beta$ for $\beta\gg 1$ 
can be written as 
a power of $\beta$: 
\begin{align}
  d_n^2((x,\beta),(x+dx,\beta)) 
  = \mathrm{const.}~\beta^q \sum_{\mu=1}^D dx_\mu^2 \quad (\beta\gg 1), 
\end{align}
where $q$ is a constant. 
On the other hand, 
the functional form of $f(\beta)$ for $\beta\gg 1$ can be determined by 
evaluating the distance in the $\beta$ direction 
from the definition~(\ref{eq:def_distance}) as follows. 
We first approximate the local equilibrium distribution in $\beta\gg 1$ 
by Gaussian. 
Then it turns out that the distance between two points 
$(x,\beta_a)$, $(x,\beta_{a+1})$ 
is a function of the ratio $\beta_a/\beta_{a+1}$ \cite{FMU2}. 
This means that the distance in the $\beta$ direction is invariant 
under scaling $\beta \rightarrow \lambda \beta$ for large $\beta$, 
and thus we obtain 
\begin{align}
  d_n^2((x,\beta),(x,\beta+d\beta)) = \mathrm{const.}~\frac{d\beta^2}{\beta^2} \quad (\beta\gg 1). 
\end{align}

Putting everything together, 
we conclude that the metric on $\bar{\calM}\times\calA$ is given by
\begin{align}
  ds^2 = l^2 \left( \frac{d\beta^2}{\beta^2} 
  + \alpha \beta^q \sum_{\mu=1}^D dx_\mu^2 \right) \quad (\beta\gg 1)
  \label{eq:metric}
\end{align}
with constants $l,\alpha,q$. 
This is an AdS metric, 
as can be seen by the coordinate transformation 
$\beta \rightarrow (\sqrt{\alpha}q z/2)^{-2/q}$: 
\begin{align}
  ds^2 = \left( \frac{2l }{q}\right)^2 \cdot
  \frac{1}{z^2}\left( dz^2 + \sum_{\mu=1}^D dx_\mu^2 \right), 
\end{align}
which is a Euclidean AdS metric in the Poincar\'{e} coordinates. 

\begin{figure}[t]
  \centering
  \includegraphics[width=4.2cm]{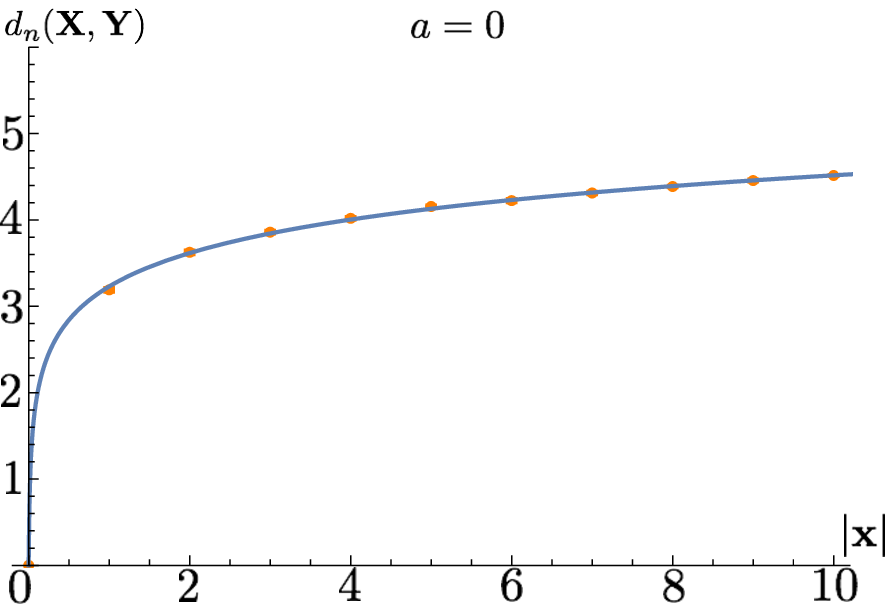}
  \includegraphics[width=4.2cm]{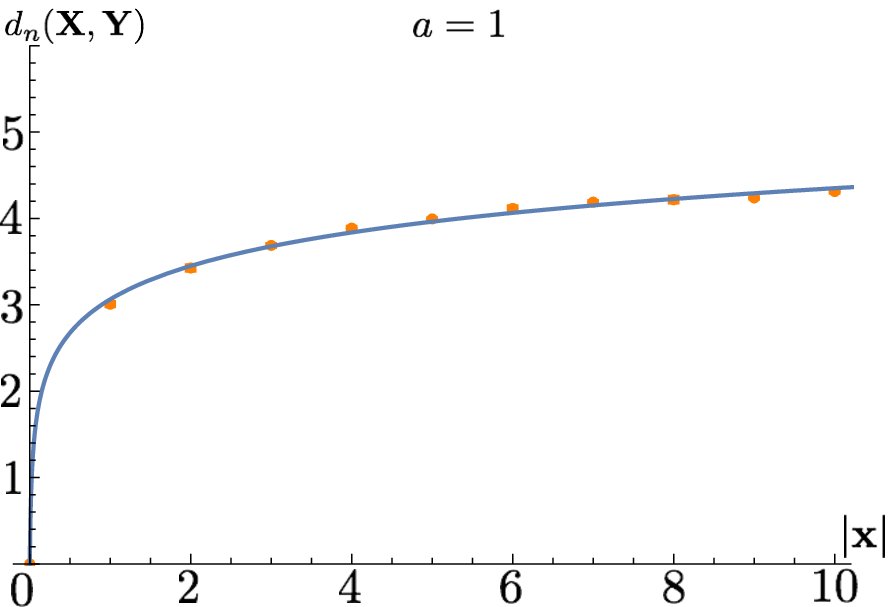}
  \includegraphics[width=4.2cm]{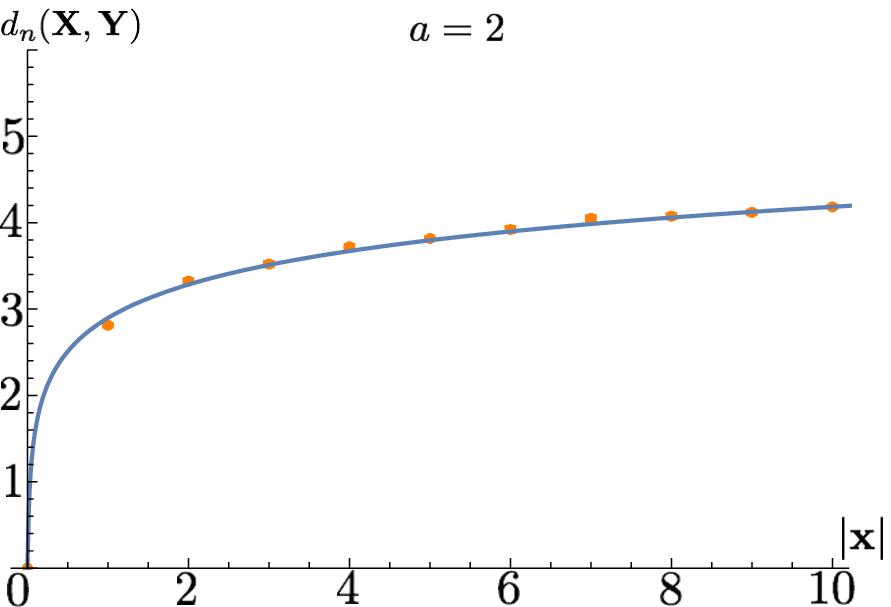}
  \caption{
    Calculated distances \cite{FMU2}. 
    The solid line is the geodesic distance
    with the fitted parameters.}
  \label{fig:adsfit_opt}
\end{figure}

We can verify this metric in the following way. 
We first numerically calculate the distance 
$d_n(X\equiv(0,\beta_a),Y\equiv(x,\beta_a))$ 
for $a=0,1,2$ and $x=1,\cdots, 10$. 
We then make a $\chi^2$ fit by using as the fitting function 
the geodesic distance calculated from the metric~(\ref{eq:metric}): 
\begin{align}
  \mathcal{I}(x,\beta_a;l,\alpha,q) \equiv 
  \frac{4l}{q} \ln \left(
  \frac{\sqrt{(q \sqrt{\alpha}|x|/4)^2 + \beta_a^{-q}} 
  + q\sqrt{\alpha}|x|/4}{\beta_a^{-q/2}} \right). 
\end{align}
We carried out the above calculations 
for a two-dimensional $(D=2)$ configuration space, 
and obtained the results shown in Fig.~\ref{fig:adsfit_opt} \cite{FMU2}. 
The parameters are determined to be
$l = 0.0404(14)$, 
$\alpha = 2.34(48) \times 10^5$, 
$q = 0.289(12)$ 
with $\sqrt{\chi^2/(30-3)} = 2.7$. 
The good agreement shows that the 
distances can be regarded as geodesic distances 
of an asymptotically Euclidean AdS metric. 

\subsection{Geometrical optimization}
\label{sec:geometric_optimization}

Our aim here is to optimize the functional form of $\beta_a = \beta(a)$ 
so that transitions in the extended direction become smooth. 
We make this optimization 
by referring to the geometry of the extended configuration space. 
Note that, since it is the parameter $a$ 
that is directly dealt with in MCMC simulations, 
we expect that the smooth transitions correspond to a flat metric 
in the extended direction 
when $a$ is used as one of the coordinates: 
$g_{\beta\beta} \, d\beta^2 = \mathrm{const.}~da^2$. 
Since the geometry of $\bar{\calM}\times\calA$ 
is asymptotically AdS (\ref{eq:metric}), 
this means that $d\beta^2/ \beta^2 \propto da^2$. 
This in turn determines the functional form of $\beta_a$ 
to be exponential in $a$, 
$\beta_a = \beta_0 R^{-a}$ ($\beta_0, R$: constants). 

\begin{figure}[bht]
  \centering
  \includegraphics[width=5.5cm]{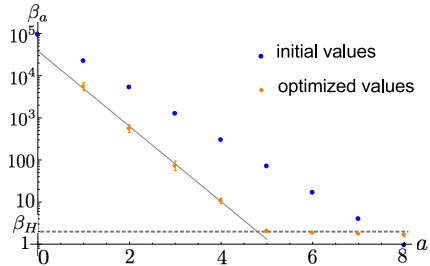}
  \caption{
    Optimized values for $\{\beta_a\}$ $(a=1,\ldots,8)$ \cite{FMU2}.
    The blue dots are the initial values, 
    and the orange dots are the resulting optimized values. 
  }
  \label{fig:beta_both}
\end{figure}

We confirmed this expectation numerically 
by gradually changing the value of $\beta_a$ 
so that the distances between different modes are minimized \cite{FMU2}. 
The result is shown in Fig.~\ref{fig:beta_both}. 
We see that the optimized form of $\beta_a$ 
certainly takes an exponential form. 

\section{Tempering parameter in the tempered Lefschetz thimble method}
\label{sec:TLTM}

The tempered Lefschetz thimble method (TLTM) 
\cite{Fukuma:2017fjq,Fukuma:2019wbv,FMU6} 
(see also \cite{FMU_Lattice2019_TLTM}) 
is an algorithm towards solving the sign problem. 
In this algorithm, by deforming the integration region 
from $\mathbb{R}^N$ to $\Sigma\subset\mathbb{C}^N$, 
we reduce the oscillatory behavior of the reweighted integrals 
that appear in the following expression: 
\begin{align}
  \langle \mathcal{O}(x) \rangle 
  = \frac{\int_\Sigma dz\,e^{-S(z)}\,\mathcal{O}(z)}
  {\int_\Sigma dz\,e^{-S(z)}}
  = \frac{\int_{\mathbb{R}^N} dx \, 
  e^{-\mathrm{Re}S(x)} e^{-i\mathrm{Im}S(x)}\mathcal{O}(x) / 
  \int_{\mathbb{R}^N} dx \, e^{-\mathrm{Re}S(x)} }
  {\int_{\mathbb{R}^N} dx \, e^{-\mathrm{Re}S(x)} e^{-i\mathrm{Im}S(x)} / 
  \int_{\mathbb{R}^N} dx \, e^{-\mathrm{Re}S(x)}}. 
\end{align} 
In Lefschetz thimble methods, 
such a deformation is made according to  
the antiholomorphic gradient flow: 
$\dot{z}^i_t = [ \partial_i S(z_t) ]^\ast$ 
with
$z^i_{t=0} = x^i$, 
where the dot denotes the derivative with respect to $t$. 
This flow equation defines a map 
from $x\in\mathbb{R}^N$ to $z=z_t(x)\in\mathbb{C}^N$, 
and a new integration surface is given by 
$\Sigma_t \equiv z_t(\mathbb{R}^N)$. 
$\Sigma_t$ approaches a union of Lefschetz thimbles 
$\{\mathcal{J}_\sigma\}$ as $t\rightarrow\infty$, 
and the integrals remain unchanged under the continuous deformation 
thanks to Cauchy's theorem. 
Each thimble $\mathcal{J}_\sigma$ has a critical point 
$z_\sigma$ (where $\partial_{z^i} S(z_\sigma)=0$), 
and configurations $z\in\mathcal{J}_\sigma$ give the same phase,
$\mathrm{Im} S(z) = \mathrm{Im} S(z_\sigma) = \mathrm{const}$. 
Thus, the oscillatory behavior of the reweighted integrals 
will get much reduced for large $t$. 
In the TLTM, 
we implement a tempering algorithm by choosing the flow time $t$ 
as the tempering parameter in order to cure the ergodicity problem 
caused by infinitely high potential barriers 
between different thimbles.

We can give a geometrical argument 
that the optimized form of flow times $t_a$ is linear in $a$ 
\cite{Fukuma:2019wbv}. 
In fact, at large $t$, 
$\mathrm{Re}S(z_t(x))$ increases exponentially as 
$\beta_t\propto e^{\mathrm{const.}~t}$. 
As in the simulated tempering, 
we expect that the optimal form of $\beta_{t_a}$ 
is an exponential function of $a$. 
Therefore, $t_a$ should be a linear function of $a$ 
(see also discussions in \cite{Alexandru:2017oyw}). 

\begin{figure}[htb]
  \centering
  \includegraphics[width=5.5cm]{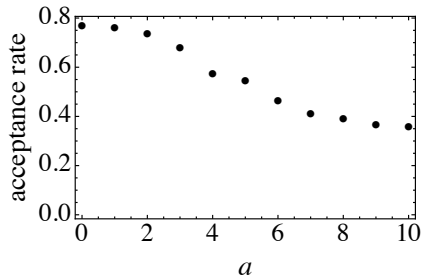}
  \caption{
    Acceptance rates in the $t_a$ direction with $\beta\mu = 8$. 
    Larger $a$ corresponds to larger $t_a$. 
  }
  \label{fig:acceptance}
\end{figure}

In the application of TLTM to the Hubbard model \cite{Fukuma:2019wbv}, 
we confirmed that this choice actually works well. 
Fig.~\ref{fig:acceptance} shows the acceptance rates 
between adjacent time slices, 
where $t_a$ is taken to be a piecewise linear function of $a$ 
with a single breakpoint. 
This choice results in the acceptance rates being roughly above 0.4. 
Most notably, the acceptance rates become constant for larger $t$ (larger $a$), 
where $\Sigma_t$ gets close to the thimbles 
and the above discussion becomes more valid. 

\section{Conclusion and outlook}
\label{sec:conclusion}

We introduced the distance between configurations, 
which quantifies the difficulty of transitions. 
We then discussed that an asymptotically AdS geometry emerges 
in the extended, coarse-grained configuration space, 
and showed that the optimization of the tempering parameter 
can be made in a simple, geometrical way. 
We further argued how to determine the optimized form 
of the tempering parameter in the tempered Lefschetz thimble method. 

As for future work, 
it should be interesting to investigate the distance in the Yang-Mills theory, 
where the coarse-graining of the configuration space can be made 
by identifying configurations with the same topological charge 
as a single configuration. 
We further would like to apply the distance to models 
whose degrees of freedom can be 
interpreted as spacetime coordinates (e.g., matrix models) \cite{FM}. 
Then the geometry of the configuration space 
directly gives that of a spacetime. 
We expect that 
this formulation gives a systematic way 
to construct a spacetime geometry from randomness 
and provides us with a way to define a quantum theory of gravity. 

A study along these lines is now in progress and will be reported elsewhere. 

\acknowledgments
The authors greatly thank the organizers of LATTICE 2019. 
They also thank Andrei Alexandru, Hiroki Hoshina, Etsuko Itou, Yoshio Kikukawa, 
Yuto Mori and Akira Onishi for useful discussions. 
This work was partially supported by JSPS KAKENHI 
(Grant Numbers 16K05321, 18J22698 and 17J08709) 
and by SPIRITS 2019 of Kyoto University (PI: M.F.).

\end{document}